\documentclass{birkjour}
 \theoremstyle{definition}
 
 \theoremstyle{remark}

 \numberwithin{equation}{section}

\begin{document}

\title{Noncommutative complex Grosse-Wulkenhaar model}

\author[Mahouton Norbert Hounkonnou]{Mahouton Norbert Hounkonnou}

\address{%
University  of Abomey-Calavi\\  International Chair in
Mathematical Physics
and Applications\\
 ICMPA-UNESCO  CHAIR\\ 072B.P.:50, Cotonou, Rep. of Benin
}

\email{norbert.hounkonnou@cipma.uac.bj with copy to hounkonnou@yahoo.fr}

\thanks{CP1079, Geometric Methods in Physics—Proceedings of the XXVII Workshop on Geometric Methods in Physics
edited by P. Kielanowski, A. Odzijewicz, M. Schlichenmaier, and T. Voronov
O 2008 American Institute of Physics 978-0-7354-0610-0/08/}
\author{Dine Ousmane Samary}
\address{University  of Abomey-Calavi\\  International Chair in
Mathematical Physics
and Applications\\
 ICMPA-UNESCO  CHAIR  \\072B.P.:50, Cotonou, Rep. of Benin}
\email{ousmanesamarydine@yahoo.fr}

\keywords{Noncommutative field theory, Moyal algebra,
Noether theorem, Ward identity operator, Energy-momentum tensor.}

\date{January 1, 2004}

\begin{abstract}
This paper  stands for an application of
 the noncommutative (NC) Noether theorem, given in our previous work
 [AIP Proc {\bf 956} (2007)  55-60], for the NC complex Grosse-Wulkenhaar model. It  provides
 with an extension of a recent work  [Physics Letters {\bf B 653} (2007) 343-345].
  The local conservation of energy-momentum tensors
 (EMTs) is recovered using improvement procedures based
 on Moyal algebraic techniques.  Broken dilatation symmetry is  discussed.
   NC gauge currents are also explicitly computed.
\end{abstract}

\begin{flushright}
ICMPA-MPA/2008/12 \\
\today
\end{flushright}
\maketitle


\section{ Introduction}

Field theory on noncommutative geometry, or noncommutative field theory (NCFT), for which there
  exists a respectable body of mathematical and physical
knowledge (see [1]-[15] and references therein),  is most often performed
over a Moyal space
${\rm I\!\!R}_\theta^D$, a deformed D-dimensional space endowed with a constant Moyal $\star-$bracket of
coordinate functions
 \begin{eqnarray}
\left[x^\mu, x^\nu\right]_{\star} = i \Theta^{\mu\nu}
\end{eqnarray}
where $\Theta$ is a $D\times D$ non-degenerate skew-symmetric matrix (which requires D even),
 usually  chosen in canonical form as
 \begin{eqnarray}
 \Theta=\theta J \mbox{ with } J=
\left(\begin{array}{ll}
0 & I_{N}\\
-I_{N} & 0
\end{array}
\right).
\end{eqnarray}
$]0, +\infty[\ni\theta $ is a square length dimensional parameter,
($[\theta] = [L]^2$).
 The corresponding product of functions is the associative, noncommutative
Moyal-Groenewold-Weyl product,
simply called hereafter {Moyal product or $\star$-product}
defined by
\begin{eqnarray}
(f\star g)(x) = {\rm m}\left\{
e^{i \frac{\Theta^{\rho\sigma}}{2}
 \partial_{\rho}\otimes \partial_{\sigma}} f(x)\otimes g(x)
 \right\}, \quad x\in {\rm I\!\!R}_\theta^D, \qquad  \forall f,g\in\mathcal{S}({\rm I\!\!R}_\theta^D)
 \end{eqnarray}
${\rm m}$ is the   ordinary multiplication of functions and
 $\mathcal{S}({\rm I\!\!R}_\theta^D)$ - the space of suitable Schwartzian functions.
 For precise definition, see [11]-[14].
  Such a noncommutative
 geometry possesses the  specific pathology to break both the Lorentz invariance by the presence of  $\Theta^{\mu\nu}$,
 as $\left[x^\mu, x^\nu\right]_{\star} = i \Theta^{\mu\nu}$ is not generally invariant under rotation, and the
 local character of the theory thanks to the property that
 \begin{eqnarray}
 (f\star g)(z)& =&  \int d^Dx d^Dy K(x,y;z)f(x)f(y)
 \end{eqnarray}
is evaluated through a two-point kernel
 \begin{eqnarray}
 K(x,y;z)=\delta(z-x)\star\delta(z-y)
 ={1\over{(2\pi)^D det\Theta}}e^{i(z-x)\Theta^{-1}(x-y)}.
 \end{eqnarray}
Besides, the uncertainty principle is expressed by
$\Delta x\Delta y\ge \theta$.  For more details on the NCFT characterization, see [11] and references therein.
Peculiar features of noncommutative field theories (NCFT)
in the Moyal $\star$-product description
generally engender energy momentum tensors (EMTs) which are not locally conserved,
not traceless in the massless situation and,
not symmetric and not gauge invariant in gauge theories
\cite{Gerh,Grims,Abu,Das,bghnk1,bghnk0}.

Noether's procedure for infinitesimal translation symmetry in
noncommutative (NC) scalar field theory has been studied by Gerhold
{\it et al.}\cite{Gerh} and by Abou-Zeid and Dorn \cite{Abu}.
According to their works, the canonical EMT appears to be symmetric
albeit not locally conserved. From algebraic techniques, it can be
obtained an improved locally conserved EMT \cite{Abu}.
Nevertheless, this improved tensor fails to be symmetric.
It is noteworthy that, in classical field theory,
a symmetric EMT is related to a locally
conserved angular momentum. Further, in NCFT, the constant tensor
$\Theta^{\mu\nu}=-i[x^\mu,x^\nu]$, characterizing the noncommutative
geometry, is not a Lorentz tensor and obviously breaks the Lorentz
invariance. Therefore, the nonlocal conservation of angular momentum
and then the asymmetry of EMT are {\it a priori} allowed.

 NC versions of Noether currents have
been worked out for translation, dilatation and local gauge
symmetries \cite{Gerh}-\cite{wess}.

So far, all these studies have been performed
before the advent of the new class of renormalizable NC
field theories (NCRFT) built on the Grosse and Wulkenhaar
$\phi^4$ scalar field theory \cite{GW,Riv}.

 In our previous work \cite{bghnk0},  an explicit formulation
 of an analogue of the Noether
theorem in Moyal NCFTs has been provided at the operator level
as follows.  Given $W$ the global canonical NC  Ward identity operator (WIop) for a
transformation, then the following statement holds:

 {\bf Theorem} (NC Noether theorem, Ben Geloun and Hounkonnou):
{\it If a NC action $S$ is invariant under
a set of transformations, then
$$W \,S = - \int d^Dx \;{\rm Tr}  \;\partial^\mu J_{\mu} = 0,$$
and there exists a globally conserved vector current $J_{\mu}$.
}

 Albeit the local conservation of the current is not
required, many facts invite us to consider this statement as actually the
best of what we can expect from the NC action
formulation. Indeed, $(i)$
 it is the NC method the closest
of usual Noether procedure;  $(ii)$
in NC space-space $\theta^{0i}=0$ geometry, there exists
an exactly conserved $D$-vector momentum $P_\mu$
for infinitesimal translation symmetry which readily
generalizes its undeformed counterpart;
$(iii)$ it can be proved that
regularization procedures improve quantities to
recover local conservation property
for many field theories and
$(iv)$ finally, at the parameter limit $\theta\to 0$
the globally conserved currents obtained through the
NC WIop method reduce to classical Noether currents.

Recently, this statement has been successfully applied to investigate
EMTs in renormalizable noncommutative
scalar field theory \cite{bghnk1}.

The purpose of this paper is to give, as another original contribution,
 an application of
 the above  NC Noether theorem in the framework of the NC complex Grosse-Wulkenhaar (GW) model,  to
 perform the local conservation of energy-momentum tensors
 (EMTs),  using improvement procedures based
 on Moyal algebraic techniques, as well as to analyze the  broken dilatation symmetry and to
  explicitly compute NC gauge currents.

\section{ Noncommutative complex GW model}
 The NC complex Grosse-Wulkenhaar Lagrangian action can be written as:
\begin{eqnarray}\label{eq:action}
S_{\star}^{\Omega}[\phi,{\bar \phi}] &=& \int d^{D}x\,\Big[\partial_{\mu}\phi\star\partial^{\mu}
{\bar \phi} +m^{2}\phi\star{\bar \phi}
+\frac{\Omega^{2}}{2}(\tilde{x}_{\mu}\phi)\star
(\tilde{x}^{\mu}{\bar \phi}) \cr
&+&\frac{\lambda}{2.4!}(\phi\star{\bar \phi}\star\phi
\star{\bar \phi}
+\phi\star{\bar \phi}\star{\bar \phi}\star\phi)\Big],
\end{eqnarray}
where { $\widetilde{x}= 2(\Theta^{-1})\,\cdot\, x$},
{ $\Theta$} breaks into diagonal blocks
 $\left(\begin{array}{ll}
0 & \theta \\
-\theta & 0
\end{array}\right)$.
{$\phi$} is a complex scalar field (with rapid decay).
 The harmonic term $\Omega$ ensures ultraviolet
 (UV)/infrared (IR)
freedom for the action implying its renormalizability,
 and such that the Lagrangian
action becomes covariant under Langmann-Szabo duality  \cite{LS},
 i.e. covariant under the symmetry:
 $\tilde{x}_{\mu}\longleftrightarrow p_{\mu}\equiv \partial_{\mu}.$
 The Lagrangian density depending explicitly
on $x^\mu$, through the field $\phi$  interaction with a harmonic
external source,  does not describe a closed system.   Furtheremore, it is not
invariant under space-time translation.  Besides,
 at the parameter limit $\theta \to 0$,
the model does not converge to the ordinary $\phi^4$ scalar field
theory due to the presence of the inverse matrix $(\Theta^{-1})$,
then causing a singularity.
 The   { $\star$}-Grosse-Wulkenhaar
{ $\phi_D^4$}  theory is renormalizable at all orders in $\lambda$.
This result has been now proved by  various methods (see \cite{Riv} and references therein).
It is a matter of algebra to recast the
Grosse-Wulkenhaar harmonic term as
\begin{eqnarray}
(\tilde{x}_{\mu}\phi)\star(\tilde{x}^{\mu}{\bar \phi})&=&
\frac{1}{4}
\left(
\tilde{x}_{\mu}\star\phi\star\tilde{x}^{\mu}
 \star{\bar \phi}
\right.
+ \tilde{x}_{\mu}\star\phi\star{\bar \phi}\star
\tilde{x}^{\mu}
\cr
& +& \phi\star\tilde{x}_{\mu}\star\tilde{x}
^{\mu}\star{\bar \phi} +\phi\star\tilde{x}_{\mu}\star{\bar \phi}
\star\tilde{x}^{\mu} )\nonumber
\end{eqnarray}
and to re-express accordingly the action (\ref{eq:action}) so that it now entirely lies
in the $\star$-algebra of fields with the advantage
to be stable under formal $\star$-algebraic computations
(such that the cyclicity of $\star$-factors under integral).
We get the equations
of motion for $\phi$ and $\bar \phi$ as follows:
  \begin{eqnarray}
\frac{\delta S_{\star}^{\Omega}}{\delta\phi}= 0 \Leftrightarrow &-&\partial_{\rho}\partial^{\rho}
{\bar \phi}+m^{2}{\bar \phi}
+\frac{\lambda}{2.4!}(2{\bar \phi}
\star\phi\star{\bar \phi}+{\bar \phi}\star{\bar \phi}\star\phi
+\phi\star{\bar \phi}\star{\bar \phi})\cr
                             &+&\frac{\Omega^{2}}{8}(2
\tilde{x}\star{\bar \phi}\star\tilde{x}+
{\bar \phi}\star\tilde{x}\star\tilde{x}
+\tilde{x}\star\tilde{x}\star{\bar \phi})=0, \nonumber
\end{eqnarray}
\begin{eqnarray}
\frac{\delta S_{\star}^{\Omega}}{\delta{\bar \phi}}= 0 \Leftrightarrow  &-&\partial_{\rho}\partial^{\rho}
\phi+m^{2}\phi
+\frac{\lambda}{2.4!}(2\phi\star{\bar \phi}\star
\phi+\phi\star\phi\star{\bar \phi}+{\bar \phi}\star\phi\star\phi)\cr
                             & +&\frac{\Omega^{2}}{8}(2
\tilde{x}\star\phi\star\tilde{x}+
\phi\star\tilde{x}\star\tilde{x}
 +\tilde{x}\star\tilde{x}\star\phi)=0.\nonumber
\end{eqnarray}
One avoids translational invariance violation for the appearance of
the coordinate $\widetilde{x}^\rho$ by considering further the
constraint ${\delta S_{\star}^{\Omega}}/{\delta\tilde{x}_{\rho}}= 0$, i.e.
 \begin{align}\label{eq:constraint}
&& \frac{\Omega^{2}}{8}(
2\phi\star\tilde{x}^{\rho}\star{\bar \phi}+2{\bar \phi}\star
\tilde{x}^{\rho}\star\phi+\phi\star{\bar \phi}\star\tilde{x}^{\rho}\cr
& &+\tilde{x}^{\rho}
\star\phi\star{\bar \phi}+\tilde{x}^{\rho}\star{\bar \phi}\star\phi
+{\bar \phi}\star\phi\star\tilde{x}^{\rho})=0.
\end{align}
From infinitesimal  translations, we can now define the global canonical
Ward identity operator for the NC complex GW model
 \begin{eqnarray}
{\mathcal W}^{\theta}_{\mu}&=&\int d^{D}x\,(\partial_{\mu}\phi\star
\frac{\delta}{\delta\phi}+\frac{\delta}{\delta\phi}\star\partial_{\mu}\phi+\partial_{\mu}{\bar \phi}\star
\frac{\delta}{\delta{\bar \phi}}+\frac{\delta}{\delta{\bar \phi}}\star\partial_{\mu}{\bar \phi}\cr
&+&\partial_{\mu}\tilde{x}_{\rho}
\star\frac{\delta}{\delta\tilde{x}_{\rho}}+\frac{\delta}{\delta\tilde{x}_{\rho}}\star\partial_{\mu}\tilde{x}_{\rho})
\end{eqnarray}
such that its action on the Lagrangian density
 \begin{eqnarray}
{\mathcal W}^{\theta}_{\mu} S_{\star}^{\Omega}\equiv -\int d^D x \;\partial^\rho\; T_{\rho\mu}^{\Omega}=0
\end{eqnarray}
gives the  canonical energy momentum tensor (EMT)
\begin{eqnarray} \label{eq:emt}
 T_{\rho\mu}^{\Omega}= \frac{1}{2}\{\partial_{\mu}\phi,\partial_{\rho}
{\bar \phi}\}_{\star}+\frac{1}{2}\{\partial_{\mu}{\bar \phi},
\partial_{\rho}\phi\}_{\star} - g_{\rho\mu}\mathcal{L}_{\star}^{\Omega},
\end{eqnarray}
 where $g_{\rho\mu}$ is the Euclidean metric,
$\mathcal{L}_{\star}^{\Omega}$ the NC Lagrangian,
$\{(\cdot),(\cdot)\}_{\star}$ the
$\star$-anticommutator.
The EMT then conserves its form comparatively to
the result of \cite{bghnk1}.
$ { T_{\rho\mu}^{\Omega}}$ is symmetric, nonlocally conserved,
and in massless theory, not traceless.
Moreover, putting the mass term to zero,
the usual improved tensor
\begin{eqnarray}
 T_{\rho\mu}^{I,\Omega}&=&T_{\rho\mu}^{\Omega ,m=o}
 +\frac{1}{6}\Big(g_{\rho\mu} \fbox{}
  - \partial _{\rho}\partial _{\mu}\Big)\{\phi ,{\bar \phi}\}_{\star}
 \nonumber
\end{eqnarray}
is not traceless too.
Let us investigate now an improvement to (\ref{eq:emt})
for the local conservation order. After some algebra,
it can be deduced
 \begin{eqnarray} \label{local}
\partial^{\rho}T_{\rho\mu}^{\Omega}&=&\partial^{\rho}T_{\rho\mu}-\frac{\Omega^{2}}{16}
\Big([\tilde{x}\star\tilde{x},\partial_{\mu}\phi\star{\bar \phi}
+\partial_{\mu}{\bar \phi}\star\phi]_{\star}+[\tilde{x}\star{\bar \phi},
\partial_{\mu}\phi\star\tilde{x}]_{\star}\cr
 &+&
[\tilde{x}\star\phi,\partial_{\mu}{\bar \phi}\star\tilde{x}]_{\star}
 +[\tilde{x}\star
\partial_{\mu}\phi,{\bar \phi}\star\tilde{x}]_{\star}
+[\tilde{x}\star\partial_{\mu}{\bar \phi},\phi\star\tilde{x}]_{\star}\Big)\cr
&-&\frac{\Omega^{2}}{8}[\tilde{x},\partial_{\mu}\phi\star\tilde{x}\star{\bar \phi}+
\partial_{\mu}{\bar \phi}\star\tilde{x}\star\phi]_{\star}\cr&=:&-\partial^\rho t_{\rho\mu}^{\Omega}.\nonumber
\end{eqnarray}
where
  \begin{eqnarray}\label{t}
\partial^{\rho}T_{\rho\mu} &=& -\frac{\lambda}{4(4!)}\Big([
\phi\star{\bar \phi},\partial_{\mu}\phi\star{\bar \phi}-
\phi\star\partial_{\mu}{\bar \phi}]_{\star}
+[{\bar \phi}\star
\phi,\partial_{\mu}{\bar \phi}\star\phi-{\bar \phi}\star
\partial_{\mu}\phi]_{\star}\cr
                            &+&\frac{1}{2}[\phi\star
{\bar \phi},\partial_{\mu}{\bar \phi}\star\phi-{\bar \phi}\star
\partial_{\mu}\phi]_{\star}
+\frac{1}{2}[{\bar \phi}\star\phi
,\partial_{\mu}\phi\star{\bar \phi}-\phi\star\partial_{\mu}
{\bar \phi}]_{\star}
                           \cr &+&\frac{1}{2}[\phi\star
\phi,[\partial_{\mu}{\bar \phi},{\bar \phi}]_{\star}]_{\star}
+
\frac{1}{2}
[{\bar \phi}\star{\bar \phi},[\partial_{\mu}\phi,\phi]_{\star}]_{\star
}\Big).
\end{eqnarray}
 $[(\cdot),(\cdot)]_{\star}= \mbox{Moyal $\star$-commutator}$.
 Thus  the "Wulkenization" \cite{Riv} process  clearly governs the
EMT improvement mechanism.   A closer look on (\ref{local}) shows that
$T_{\rho\mu}^{\Omega}$
is globally conserved, as in NCFT the
 $\star$-commutators under integral cancel. For physical interpretation,
let us consider the space-space NCFT determined by $\Theta^{0i}=0$. Then one can readily
prove that there exists a  conserved
$D$-vector  momentum $ P_\mu^{\Omega}$, namely $\partial^0 P_\mu^{\Omega}=
\partial^0 \int d^{D-1} x\, T_{0\mu}^{\Omega}=0$. Such a vector conserved
quantity is also observed in the {\it naive} (unrenormalizable) NC
scalar field \cite{Gerh} as well as in the NC real GW model \cite{bghnk1}.
 Besides, it turns out that, displaying  the same  tedious algebraic apparatus as in
  \cite{bghnk1}, a correction term can be provided to get a new
    locally conserved albeit nonsymmetric EMT,
    $ \hat{T}_{\rho\mu}^{I,\Omega}=  {T}_{\rho\mu}^{I,\Omega} +
    t_{\rho\mu}^{\Omega}$, with nonvanishing trace for m = 0 recalling that the
    theory is not scale invariant.
  Furtheremore, one can work out a  symmetric
locally conserved EMT  through the ordinary Belifante trick (see \cite{bghnk1} and references therein),
 defining the tensor
 $\chi_{\sigma\rho\mu}$ such that
   $$\hat{T}_{\rho\mu}^{\Omega s} =\hat{T}_{\rho\mu}^{\Omega}  +
   \partial^{ \sigma} \chi_{\sigma\rho\mu},
\quad \chi_{\sigma\rho\mu} = -\chi_{\rho\sigma\mu}.$$
The  underlying Belifante type partial differential equation
 \begin{eqnarray}
 \hat{T}_{\rho\mu}^{\Omega}  - \hat{T}_{\mu\rho}^{\Omega} = \partial ^{\sigma}
(\chi_{\sigma\mu\rho} -\chi_{\sigma\rho\mu}  )=:
\partial ^{\sigma} \chi_{\sigma[\mu\rho]}\nonumber
\end{eqnarray}
 is  less comfortable than the
one worked out by Abou-Zeid and Dorn \cite{Abu}. More detailed consideration of various
 properties of EMTS will appear elsewhere.

 Let us now better scrutinize  the nonlocal conservation
of the canonical massless EMT. It  obviously induces a dilatation
symmetry breaking. In addition,
even if an improved locally conserved EMT is provided,
the scale invariance is no longer valid and predictable
since the evidence of a nonvanishing trace of this improved EMT.
Both these arguments on dilatation symmetry breaking
are valid for massless Grosse and Wulkenhaar model. Indeed,
defining
infinitesimal dilatation generators   and the corresponding global symmetrized Wlop
 $W_{D,\epsilon}^{\theta}$, respectively,
 \begin{eqnarray}
\delta_{\epsilon}\tilde{x}_{\mu}&=&(1+\epsilon)\tilde{x}_{\mu},\quad
 \delta_{1,\epsilon}
\phi=\epsilon D_{1}\phi,\quad \delta_{2,\epsilon}\phi=\epsilon
D_{2}\phi,
  \cr
D_{1}(.)&=&(1+x^{\mu}\star\partial_{\mu})(.),\quad D_{2}(.)=((.)+\partial_{\mu}(.)\star x^{\mu})
\end{eqnarray}
\begin{eqnarray}
 W_{D,\epsilon}^{\theta}(.)&=&\int d^{D}x\Big\{\frac{\epsilon}{4}\Big[D_{1}\phi\star\frac{\delta(.)}{\delta \phi}
  +\frac{\delta(.)}{\delta \phi}\star D_{1}\phi+D_{2}\phi\star\frac{\delta(.)}{\delta \phi}
 +\frac{\delta(.)}{\delta \phi}\star D_{2}\phi\cr&+&(\phi\longleftrightarrow {\bar \phi})\Big]\
 +\frac{1}{2}\Big[\delta_{\epsilon}\tilde{x}_{\rho}\star\frac{\delta(.)}{\partial\tilde{x}_{\rho}}
+\frac{\delta(.)}{\delta\tilde{x}_{\rho}}\star\delta_{\epsilon}\tilde{x}_{\rho}\Big]\Big\} \nonumber
\end{eqnarray}
 so that
$\frac{\partial}{\partial \epsilon} W_{D,\epsilon}^{\theta}
(S^{\Omega}_{\star})=-\int d^{D}x(\partial^{\rho}\mathcal{D}_{\rho}^{\Omega}+ B^{\Omega}_{\star})$
with
 $\mathcal{D}_{\rho}^{\Omega}$ the dilatation current given by
 $\mathcal{D}_{\rho}^{\Omega}=\frac{1}{2}\{x^{\mu},\hat{T}_{\rho\mu}^{I,\Omega}\}_{\star}$,
 the breaking quantity  $B_{\star}^{\Omega}$ reveals to depend both
  on the nonvanishing trace of the local
  conservation improving tensor $t_{\rho\mu}^\Omega$ through $\hat{T}_{\mu}^{I,\Omega,\mu}$
  and on the GW term as follows:
\begin{eqnarray}
&& B_{\star}^{\Omega}= -\hat{T}_{\mu}^{I,\Omega,\mu}-\frac{1}{2}\{x^{\mu},
 \frac{\lambda}{2.(4!)}[\partial_{\mu}\phi\star(2{\bar \phi}\star\phi\star{\bar \phi}
+\{\bar \phi\star{\bar \phi},\phi\}_{\star})
\cr
&&+\partial_{\mu}{\bar \phi}(2\phi\star{\bar \phi}\star
\phi
+\{\bar \phi,\phi\star\phi\}_\star)
-\frac{1}{2}\partial_{\mu}(\{\phi\star{\bar \phi}\star
\phi,{\bar \phi}\}_\star)\cr
&&+\frac{1}{4}\partial_{\mu}(\{\phi\star{\bar \phi},{\bar \phi}
\star\phi\}_\star
+\{\phi\star\phi,{\bar \phi}\star{\bar \phi}\}_\star)]+\frac{\Omega^{2}}{8}[-\partial_{\mu}(\{\tilde{x}\star\phi,\tilde{x}\star{\bar \phi}\}_\star\cr
 &&+\frac{1}{2}\{\tilde{x}\star\{\bar \phi,\phi\}_\star,\tilde{x}\}_\star)
 +\partial_{\mu}\phi\star(2
\tilde{x}\star{\bar \phi}\star\tilde{x}
+\{\bar \phi,\tilde{x}\star\tilde{x}\}_\star)\cr
&&+\partial_{\mu}{\bar \phi}\star(2
\tilde{x}\star\phi\star\tilde{x}
+\{\phi,\tilde{x}\star\tilde{x}\}_\star)]\}_{\star}
\end{eqnarray}

 Finally,  let   ${ U_\star(1)}$ be the NC gauge group generated by elements $U\in { U_\star(1)}$:
  \begin{eqnarray}
 e^{i\alpha}_{\star}=1 + i\alpha +(i^2/2!) \alpha\star\alpha
 + (i^3/3!) \alpha\star\alpha \star\alpha +\dots,
 \alpha \in C^\infty({\rm I\!\!R})
  \end{eqnarray}
 Then, defining the left group action and infinitesimal transformations
  \begin{eqnarray}
   \phi \longmapsto \phi ^\prime = U_\star\phi;\quad
  \delta _\alpha \phi = i\alpha\star\phi;\quad
  \bar\phi \longmapsto \bar\phi ^\prime = \bar\phi U_\star^\dag;\quad
  \delta _\alpha \bar\phi = -i\bar\phi\star \alpha,
   \end{eqnarray}
   and  the infinitesimal action variation
   \begin{eqnarray}
   \delta S = \int \left\{ \partial_\mu (\delta \phi \star \partial^\mu \bar \phi)
   + \partial_\mu (\partial^\mu \phi \star\delta \bar\phi)
   +
    \delta \bar\phi\star(\mathcal{EL}(\phi)) +
     (\mathcal{EL}(\bar\phi))\star \delta \phi   \right\}
    \end{eqnarray}
      where  $\mathcal{EL}(\phi)$ (resp. $\mathcal{EL}(\bar\phi)$) represents
  the  Euler-Lagrange equations of motion for $\phi$ (resp. for $\bar\phi$),
leads to express $U_\star(1)$ gauge WIop  in the form
\begin{eqnarray}
W^G (\cdot)=
 \int  d^D y \frac{   \delta }{\delta \alpha(y)}
 \int  d^D x
\left[
 \frac{\delta (\cdot)}{\delta \bar \phi} \star \delta \phi+
 \delta \bar\phi\star  \frac{\delta (\cdot)}{\delta  \phi}\right] \end{eqnarray}
such that $W^G (S) =   \int  d^D y\;\, \partial_\mu J^\mu$ yields
the suitable NC  $U_\star(1)$ gauge currents of the GW model as follows:
\begin{eqnarray}
J^\mu = i\left(\phi \star \partial^\mu \bar\phi- \partial^\mu \phi \star\bar\phi\right).
\end{eqnarray}

 \section{Conclusion}
This work has proved that the WIop formulation of the NC Noether theorem
given in \cite{bghnk0} is one of the keystones in Moyal field theory analysis.
 Its successful application here allowed to investigate Neither currents for NC complex  GW model.
 Symmetry properties (translation, dilatation and gauge symmetries) have been  studied. EMTs are regularizable by
 usual methods.
  $U_\star(1)$ gauge currents  have been explicitly computed.


\subsection*{Acknowledgment}
This work is partially supported by the ICTP through the
OEA-ICMPA-Prj-15. The ICMPA is in partnership with
the Daniel Iagolnitzer Foundation (DIF), France.
The authors thank Dr J. Ben Geloun for fruitful discussions.
MNH expresses his gratefulness to Professor A. Odzijewicz and all his staff for their hospitality
and the good organization of the Workshops in Geometric Methods in Physics.

\end{document}